\documentclass{amsart} 
\usepackage[latin1]{inputenc}
\usepackage[swedish,english]{babel} 
\usepackage{amssymb}
\usepackage{color}
\usepackage{ifpdf} 
\ifpdf
\usepackage[pdftex]{graphicx} 
\usepackage{epstopdf} 
\else
\usepackage[dvips]{graphicx} 
\fi 
\usepackage{caption} 
\usepackage{multirow}
\usepackage[pdftitle = {X-ray laser diffractions on GPU clusters},
  pdfauthor = {Ekeberg/Engblom/Liu},
  pdffitwindow = true,
  breaklinks = true,
  colorlinks = true,
  urlcolor = blue,
  linkcolor = red,
  citecolor = red,
  anchorcolor = red]{hyperref}
\usepackage[numbers,sort]{natbib} 

\usepackage{algorithm}
\usepackage{algpseudocode}
\usepackage{algpascal}

\usepackage{subfig} 
\usepackage{listings}
\usepackage{subfig}

\newcommand{\new}[1]{#1}
	
\newcommand{\Mpix}{M_{\mbox{{\tiny pix}}}}
\newcommand{\Mrot}{M_{\mbox{{\tiny rot}}}} 
\newcommand{\Mdata}{M_{\mbox{{\tiny data}}}}
\newcommand{\Mgrid}{M_{\mbox{{\tiny grid}}}}

\newcommand{\Mchunks}{M_{\mbox{{\tiny chunks}}}}
\newcommand{\Prob}{\mathbf{P}} 

\newcommand{\Wgrid}{\mathbb{W}}
	
\newcommand{\Lkernel}{\texttt{<<<}}
\newcommand{\Rkernel}{\texttt{>>>}}
\newcommand{\Bcast}{\mathbf{MPI\_Bcast}}

\newcommand{\Reduce}{\mathbf{MPI\_AllReduce}}
\newcommand{\Global}{\mathbf{\_\_global\_\_}} 
\newcommand{\void}{\mathbf{ void }}
	
\numberwithin{equation}{section} 
\numberwithin{figure}{section}
\numberwithin{table}{section}
	
\begin{document}

\title[X-ray laser diffractions on GPU clusters]{Machine learning for
  ultrafast X-ray diffraction patterns on large-scale GPU clusters}
	
\author[T. Ekeberg]{Tomas Ekeberg}
	
\author[S. Engblom]{Stefan Engblom} 
	
\author[J. Liu]{Jing Liu}
	
\address[S. Engblom \and J.~Liu]{Division of Scientific Computing,
  Department of Information Technology, Uppsala university, SE-751 05
  Uppsala, Sweden.}
	
\urladdr[S. Engblom]{\url{http://user.it.uu.se/~stefane}}
\email{stefane, jing.liu@it.uu.se}
	
\address[T. Ekeberg \and J.~Liu]{Laboratory of Molecular Biophysics,
  Department of Cell and Molecular Biology, Uppsala university, SE-751
  24 Uppsala, Sweden.}
	
\email{tomas.ekeberg, jing.liu@icm.uu.se}
	
\thanks{Corresponding author: S. Engblom, telephone +46-18-471 27 54, fax
+46-18-51 19 25.}

\date{\today}
	
	
\selectlanguage{english}
	



\begin{abstract}

The classical method of determining the atomic structure of complex
molecules by analyzing diffraction patterns is currently undergoing
drastic developments. Modern techniques for producing extremely bright
and coherent X-ray lasers allow a beam of streaming particles to be
intercepted and hit by an ultrashort high energy X-ray beam. Through
machine learning methods the data thus collected can be transformed
into a three-dimensional volumetric intensity map of the particle
itself. The computational complexity associated with this problem is
very high such that clusters of data parallel accelerators are
required.

We have implemented a distributed and highly efficient algorithm for
inversion of large collections of diffraction patterns targeting
clusters of hundreds of GPUs. With the expected enormous amount of
diffraction data to be produced in the foreseeable future, this is the
required scale to approach real time processing of data at the beam
site. Using both real and synthetic data we look at the scaling
properties of the application and discuss the overall computational
viability of this exciting and novel imaging technique.

\end{abstract}

\maketitle


\section{Introduction}
\label{sec:intro}

X-ray crystallography is currently the most successful method for
protein structure determination. A limitation is that this method
requires high-quality crystals of the sample protein. This is
particularly problematic for membrane proteins which are notoriously
hard to crystallize. This class of proteins contains about 20--30\% of
all proteins and are targeted by 50\% of modern drugs; still they
comprise less than 0.1\% of the known protein structures.

The recent construction of free-electron lasers (FEL) has the
potential to revolutionize structural biology by allowing structure
determination without the need for crystallization. FEL pulses are
intense enough that an interpretable diffraction signal can be
recorded from single proteins or viruses. Also, the pulses are short
enough to outrun the radiation damage to the particle and the
scattered data thus represents the intact particle even though the
extreme intensity will destroy the sample within picoseconds.

Since the diffraction data frames are collected one at a time and the
extremely intense X-ray pulse destroys the samples, it is impossible
to collect multiple exposures of the same particle. However, just like
in crystallography, we can use the fact that many biological particles
exist in identical copies. Data collected from many identical
particles can thus be treated as if they come from the same particle.
	
In this scheme, particles are injected into the stream of X-ray pulses
and intercepted in random orientation \cite{xrayPotential}. A
diffraction pattern represents a curved two-dimensional slice through
the modulus of the Fourier transform of the electron density of the
particle. Since the particles are assumed identical, the patterns will
correspond to different slices through the same Fourier density. If
the unknown orientations can be recovered, the diffraction patterns
can thus be assembled to the complete three-dimensional
Fourier-intensity of the particle.
	
As opposed to in crystallography, the orientation of each particle is
not directly measurable. Instead, the orientations are recovered by
maximizing the fit between the individual diffraction
patterns. Several algorithms for solving this problem have been
proposed \cite{EMC, Fung2008}. The most successful of these is the
Expansion Maximization Compression (EMC) method \cite{EMC} which has
been verified experimentally using artificial samples \cite{EMC2}, and
which was recently used for the reconstruction of the giant Mimivirus
\cite{TomasPhD}.
	
The LINAC Coherent Light Source (LCLS) \cite{Emma2010} has a
repetition rate of 120 Hz and a sustained hit-ratio of 20\% has been
achieved reproducibly. This corresponds to 1 million diffraction
patterns in a single 12 hour shift or about 4 TB of data. The European
XFEL is becoming operational in 2016 and will have a repetition rate
of 27,000 Hz \cite{Schneidmiller2011}.

The 3D-alignment algorithms are computationally very demanding, yet
high data volumes are fundamental for achieving high resolution and to
balance the low photon signal when studying smaller objects. In the
light of the above developments there is an imminent need for a
massively parallel implementation of the EMC algorithm to keep up with
the increased data rates and the increasing problem size.

Based on previous experience with an implementation for smaller
heterogeneous GPU-computers \cite{EMCmGPU_proceeding}, in this paper
we present a working fully distributed implementation targeting
large-scale clusters of hundreds of GPU computers. In an effort to
prepare for the increasing data rates we ensure in our implementation
that data can be effectively and flexibly distributed. We also devise
a kind of \emph{adaptive} iteration which allows computational
resources to be used in proportion to the resolution of the final
reconstruction. Similar techniques, we argue, will be required when
handling streaming data at the beam site.

An overview of the XFEL imaging setup and the associated computational
methodology is found in \S\ref{sec:ML}. Our data parallel and fully
distributed implementation is discussed in some detail in
\S\ref{sec:parallel}. Performance results on clusters of up to 100
GPUs, reaching up to and beyond 4 TFLOPS, are presented in
\S\ref{sec:experiments}, and a concluding discussion is found in
\S\ref{sec:conclusions}.


\section{X-ray laser diffraction and Maximum Likelihood imaging} 
\label{sec:ML}

In this section we summarize the experimental setup and the principles
behind 3D imaging with XFELs. The associated data analysis is
formulated as a hidden variable Maximum Likelihood problem which can
be handled by the Expectation-Maximization algorithm. We also describe
the current `best practice' in designing a working such
algorithm. Although we certainly expect the methodology to develop
further, it seems reasonable to believe that our implementation, or at
least a very similar one, will be used extensively when modern XFEL
facilities are increasingly being put to use.

\subsection{Ultrafast X-ray diffraction patterns}

\begin{figure}[!htb] 
\centering
\includegraphics[width=0.9\textwidth]{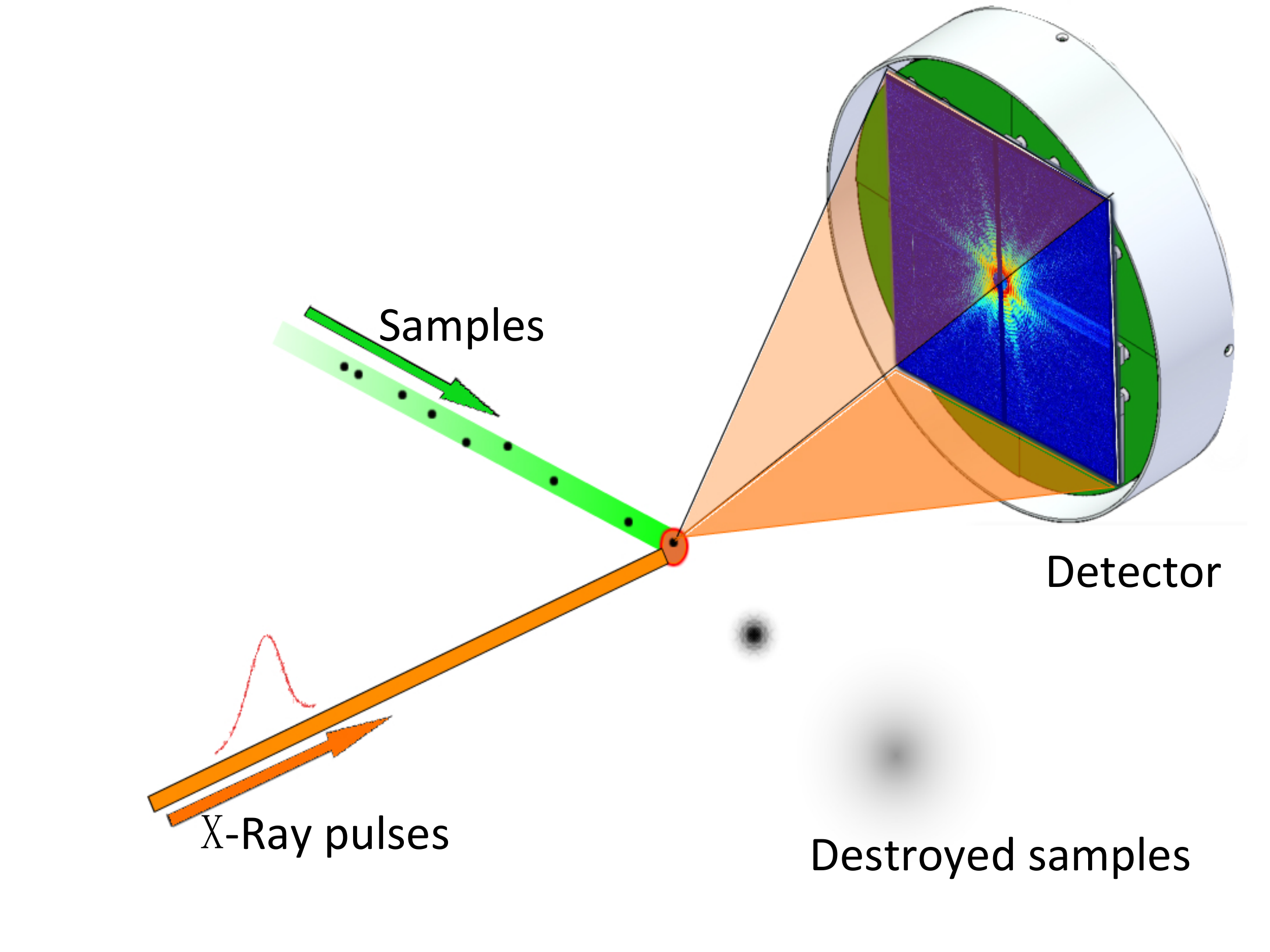} 
\caption{The principal setup in X-ray laser imaging. A stream of
  samples of biomolecules is injected and intercepts with extremely
  bright and very short pulses of X-ray lasers. While this immediately
  turns the samples into a plasma, the data collected by the photon
  count detector comes from a practically undamaged object.}
\label{fig:setup}
\end{figure}

The schematics of collecting data by XFELs is depicted graphically in
Figure~\ref{fig:setup}. An inflow of samples of biomolecules is
intercepted by an X-ray laser pulse resulting in a collection of
diffraction images. We denote the raw data output from this procedure
by $K = (K_{k})_{k = 1}^{\Mdata}$; this is a collection of
\emph{frames}, each containing measured photon counts. The detector is
discrete and hence for the $k$th frame, $K_{k} =
(K_{ik})_{i=1}^{\Mpix}$, where $\Mpix = 2^{20} = 1024 \times 1024$ is
a typical resolution. Some pixel counts near the center are missing or
may reach saturation as a result of inherent physical limitations with
the experimental procedure.

Assuming ideally that the stream of samples consists of identical
copies of a single physical object with an electron density $O$,
diffraction theory \cite{ultrafastXray} gives that each frame $K_{k}$
is given by a certain slice, an \emph{Ewald sphere}, of the 3D Fourier
transform $W$ of the real space object $O$.

With a sufficiently large collection of frames an estimate $\hat{W}$
of $W$ is first determined. Notably, this estimate lacks information
about the Fourier \emph{phases} since it is based on photon count data
only. The final step is therefore a \emph{phase retrieval
  procedure}~\cite{Chapman06,Shapiro25102005}, after which an estimate
of the real space object can be obtained.

\subsection{Maximum Likelihood estimation via Expectation-Maximization}
\label{subsec:EM}

With i.i.d.~frames $K = (K_{k})_{k = 1}^{\Mdata}$, the Maximum
Likelihood estimator is given by
\begin{align} 
\hat{W} &= \arg_{W} \max \;
\Mdata^{-1}\sum_{k = 1}^{\Mdata} \log \Prob(K_{k}| W),
\end{align} 
that is, for some probabilistic intensity model, maximizing the
likelihood of the obtained photon count data when presented with the
recorded data. The problem is incomplete for two reasons. Firstly, the
true \emph{rotation} $R_{k}$ of the object in measurement $K_{k}$ is
unknown and consequently the frame cannot be directly associated with
a definite Ewald sphere. Secondly, the energy of the X-ray pulse, the
\emph{photon fluence} $\phi_{k}$, which hits the sample is also an
unknown variable.


Besides the problem of hidden data, the overall signal to noise ratio
is very small and implies a grand computational challenge since to
counteract this the processed data volume has to be large. With input
data corrupted by high noise levels, measures has to be taken in order
to design a robust and useful algorithm \cite{EurXFEL}.

The \emph{Expectation-Maximization} (EM) algorithm~\cite{EM} aims at
producing likelihood estimates with hidden data in a constructive
way. The basic procedure of alternating steps of \textit{(i)}
assigning probabilities to the hidden states, and \textit{(ii)}
maximum likelihood estimates of the parameters of the model can be
shown under broad conditions to be at least a \emph{descent step} of
the full likelihood \cite{EMview}. Notably, in step \textit{(i)} the
model parameters are kept fixed, while in \textit{(ii)} the estimated
probabilities of the hidden states (the \emph{responsibilities} using
EM terminology) from step \textit{(i)} are assumed known.

We now introduce some notation. Firstly, the rotational space is
discretized by $(R_{j})_{j=1}^{\Mrot}$. Since this is generally a
non-uniform discretization we denote by $w_{j}$ the prior weight for
the $j$th rotation, normalized such that $\sum_{j} w_{j} = 1$. In
other words, selecting $R_{j}$ with probability $w_{j}$ implies a
practically uniform sampling of the rotational space. The intensity
space is similarly discretized by the set of points
$(q_{i})_{i=1}^{\Mpix}$ such that using this coordinate system the
unknown Fourier intensity at position $R_{j} q_{i}$ can be denoted by
$W_{ij}$. Finally, we denote by $\phi_{jk}$ the intensity of the beam
that produced data frame $k$, \emph{given} that the object was rotated
according to $R_{j}$.

An early EM algorithm for solving the problem was devised in
\cite{EMC} using the assumption that the signal is Poissonian. This
assumption has the benefit of producing an essentially parameter-free
algorithm but neglects other potential sources of noise. In practice a
Gaussian model has been more successful. More precisely we assume that
the measured intensity of the $i$th pixel in the $k$th measurement,
when scaled with the photon fluence, is Gaussian around the unknown
Fourier intensity $W_{ij}$ ~\cite{EMC2, TomasPhD}.
\begin{align}
\label{eq:gaussian_model} 
\log \Prob(K_{ik} = \kappa|W_{ij},R_{j},\phi_{jk})
&\propto -\frac{(\kappa/\phi_{jk}-W_{ij})^{2}}{2\sigma^{2}} =: Q_{ijk}(W,\phi),
\end{align} 
with $\sigma$ a noise parameter which is kept at conservative values
or decreases slightly as the iteration proceeds. Summing over $i$ we
get the joint log-likelihood function,
\begin{align}
\label{eq:jll}
Q_{jk}(W,\phi) :=
\sum_{i = 1}^{\Mpix} Q_{ijk}(W,\phi), 
\end{align} 
that is, the logarithm of the probability of observing frame $K_{k}$,
given rotation $R_{j}$ and fluence $\phi_{jk}$. Integrating this over
the space of rotations we get
\begin{align}
\nonumber 
P_{jk}^{(n+1)} &= 
P_{jk}^{(n+1)}(W^{(n)},\phi^{(n)}) :=
\Prob(R_{j}|K_{k},W^{(n)},\phi^{(n)}) \\ 
\label{eq:EstepG} 
&=
\frac{w_{j}T_{jk}(W^{(n)},\phi^{(n)})} {\sum_{j' =
1}^{\Mrot}w_{j'}T_{j'k}(W^{(n)},\phi^{(n)})}, 
\qquad
\mbox{(\textbf{``E-step''})} 
\end{align} 
in terms of $T_{jk}(W,\phi) \equiv \exp(Q_{jk}(W,\phi))$. Some care is
required when evaluating \eqref{eq:EstepG} to avoid finite precision
effects.

Although there is no explicit maximum likelihood formula for computing
$(W,\phi)$ given $P$, the following fix-point iteration for the normal
equations has been proposed \cite{EMC2},
\begin{align} 
\label{eq:MstepG} W^{(n+1)}_{ij} &=
\frac{\sum_{k = 1}^{\Mdata} P_{jk}^{(n+1)}K_{ik}/\phi_{jk}^{(n)}} {\sum_{k =
1}^{\Mdata}P_{jk}^{(n+1)}}, \\ 
\label{eq:MstepG_S} 
\phi_{jk}^ {(n+1)} &=
\frac{\sum_{i=1}^{\Mpix}K_{ik}^2} {\sum_{i=1}^{\Mpix}W_{ij}^{(n)} K_{ik}}.
\qquad \mbox{(\textbf{``M-step''})} 
\end{align} 
When the EM-iteration is understood as a descent step of the full
likelihood function, this approach of using `partial steps' can be
justified \cite{EMview}.

The rotations $(R_{j})_{j = 1}^{\Mrot}$ and the corresponding prior
weights $w_{j}$ must be found through some kind of discretization
procedure. The suggestion in \cite{EMC} is to use the fact that
quaternions encode rotations; any rotation can be identified as a
point on a 4D sphere which can hence be discretized. A suitable
geometric object for this purpose is the 600-cell (or
\emph{hexacosichoron}) which is a 4D convex regular 4-polytope whose
boundary is composed of 600 tetrahedra. At even larger values of
$\Mrot$ one further uses the \emph{fcc}-cell in which each
tetrahedron is uniformly divided $d$ times into $\{1,4,10,20,35,
...\}$ smaller tetrahedra. This implies the relation \cite[Appendix
  C]{EMC}
\begin{align} 
\label{eq:Mrot} 
\Mrot(d) &= 10 \cdot
(5d^{3}+d) \\
\nonumber 
&= [6\,300,10\,860,25\,680,50\,100,86\,520] 
\quad \mbox{for } d = [5,6,8,10,12].
\end{align} 
In \S\ref{subsec:adaptiveEMC} below we make an active use of this
discretization by increasing $d$ adaptively whenever the increase of
likelihood goes below some predefined threshold.

\subsection{EM with compression steps: the EMC} 
\label{subsec:EMC}

A problem with the EM-iteration defined by \eqref{eq:EstepG} and
\eqref{eq:MstepG}--\eqref{eq:MstepG_S} is that averages are computed
in discrete space while data is continuous. There are many pairs
$(i,j)$ such that $R_{j} q_{i}$ are very close, but in the M-step
\eqref{eq:MstepG} they will be exchanging information with disjoint or
nearly disjoint sets of frames. If the end result is to be understood
as a continuous object some kind of smoothing procedure has to be
devised.

A straightforward way to achieve this is to add
\emph{expansion/compression}-steps. The purpose of the latter step is
to compress (average/smooth) the representation into, say, a Cartesian
representation with a uniform spatial resolution. The expansion step
is the inverse of this operation and takes us back to the working
description in $R_{j} q_{i}$-space. The combination of the average
\eqref{eq:MstepG} in the M-step and a compression step then ensures
that nearby pixels and rotations have exchanged information with
overlapping sets of frames.

Let interpolation weights $f$ and interpolation abscissas $(p_{l})_{l =
1}^{\Mgrid}$ be defined such that for $g$ some smooth function, 
\begin{align}
g(q) &\approx \sum_{l = 1}^{\Mgrid} f(p_{l}-q) g(p_{l}). 
\end{align} 
An expansion operator can now be defined, 
\begin{align} 
\label{eq:estep} 
W_{ij} &=
\sum_{l = 1}^{\Mgrid} f(p_{l}-R_{j}q_{i}) \Wgrid_{l}, 
\qquad \mbox{(\textbf{``e-step''})}
\end{align} 
which maps values from a grid $\Wgrid_{l} := W(p_{l})$ into the
working description $W_{ij} = W(R_{j}q_{i})$. Similarly, a suggestion
for the compression operator is given by \cite{EMC},
\begin{align} 
\label{eq:cstep} 
\Wgrid_{l} &= \frac{\sum_{i = 1}^{\Mpix} \sum_{j= 1}^{\Mrot} f(p_{l}-R_{j}q_{i})
 W_{ij}}{\sum_{i = 1}^{\Mpix} \sum_{j =1}^{\Mrot} f(p_{l}-R_{j}q_{i})}. 
\qquad \mbox{(\textbf{``c-step''})} 
\end{align}
It should be noted that, whereas \eqref{eq:estep} is a consistent
interpolation, \eqref{eq:cstep} rather falls under the framework of
\emph{Inverse distance weighting}. Furthermore, in the implementation
discussed here, the M- and the c-steps are intertwined in that the
normalization is deferred until \emph{after} $\Wgrid$ has been
obtained.  Hence we compute (compare \eqref{eq:MstepG})
\begin{align} 
\label{eq:cMstep1}
W^{(n+1)}_{ij} &= \sum_{k = 1}^{\Mdata} P_{jk}^{(n+1)}K_{ik}/\phi_{jk}^{(n)}. \\
\intertext{The c-step is then computed as} 
\label{eq:cMstep2} 
\Wgrid_{l}^{(n+1)}
&= \frac{\sum_{i = 1}^{\Mpix} \sum_{j = 1}^{\Mrot} f(p_{l}-R_{j}q_{i})
W_{ij}^{(n+1)}}{\sum_{i = 1}^{\Mpix} 
\sum_{j = 1}^{\Mrot} f(p_{l}-R_{j}q_{i}) \sum_{k=1}^{\Mdata}P_{jk}^{(n+1)}}. 
\end{align}

In Algorithm~\ref{alg:EMC} a summary of the algorithm which will be
considered here is given. As we shall next see, the algorithm has a
distinct data parallel character and can be distributed efficiently.

\begin{algorithm} 
\begin{algorithmic} 
\State
\State \textbf{Input:} \textnormal{Initial guess of the 3D intensity
  distribution $\Wgrid^{(0)}$ of the object on the grid $(p_{l})_{l =
    1}^{\Mgrid}$, and an initial estimate of the rotational
  probabilities $P^{(0)}$.}
\State \textbf{Output:}
\textnormal{Improved image $\Wgrid$ and probabilities $P$.} 
\end{algorithmic}
\begin{algorithmic}[1] 
\Repeat \State $n = 0,1,\ldots$ 
\State $W^{(n)} := e \circ \Wgrid^{(n)}$.
\Comment{Expansion step, \eqref{eq:estep}.}
\State $P^{(n+1)} := E \circ P^{(n)}$. 
\Comment{Expectation step, \eqref{eq:EstepG}.}
\State $[W^{(n+1)},\Wgrid^{(n+1)}] := cM \circ W^{(n)}$.
\Comment{Combined Maximization and compression, 
  \eqref{eq:cMstep1} and \eqref{eq:cMstep2}.}
\Until{\mbox{change in either $\Wgrid$ or likelihood is small enough}}
\end{algorithmic} 
\caption{The EMC algorithm.} 
\label{alg:EMC} 
\end{algorithm}



\section{Parallelization in GPU clusters}
\label{sec:parallel}

By inspection the algorithm under consideration is computationally
intensive as it is composed mainly of blocks of nonlinear matrix
operations. This implies that it can be expected to perform well on
modern data parallel accelerators in general and on GPUs in
particular. The most common approach to distributed GPU computing is
to transfer data via an MPI-layer and use CUDA at the computational
nodes. This master-slave approach has been employed successfully in
other multi-GPU applications \cite{jacobsen2010mpi,
  pennycook2011performance}, and is also our approach.

In this section we take a bottom-up approach and first discuss the
single-node data parallelism and then extend the same scheme to a
distributed environment. We next devise a fully distributed approach
in which very large diffraction datasets may be considered, and we
finally also develop a simple but efficient multiresolution type
adaptivity.

\subsection{Single-node data parallelism}
\label{subsec:singleGPU}

In an implementation targeting a single GPU-node, data must be shared
between the cores of the GPU. \new{Currently, the typical resolution
  is to reconstruct a $64\times 64\times 64$ or a $128\times 128
  \times 128$ intensity model using about 1000 diffraction patterns
  (see Table~\ref{tab:dataDe}). Given the complexity of the steps of
  the algorithm, the E-step \eqref{eq:EstepG} stands out as the most
  expensive part.} Recall that the E-step estimates likelihoods
$P_{jk}$ for rotations $R_{j}$ given data frames $K_{k}$, and it
therefore makes sense to distribute the resulting probability matrix
$P$.  With a usually quite large rotational space, the algorithm
requires around 1.5 GB of memory to store the $\Mdata \times \Mrot$
matrix $P$ even for a low resolution reconstruction. Such a matrix can
be distributed by dividing the rotational space or by distributing the
images themselves. In the single-node implementation we choose not to
distribute by images since $\Mdata$ is usually much smaller than
$\Mrot$.

Our single-node EMC was implemented using CUDA with C/C++ wrappers and
the implementation closely follows the logic in
Algorithm~\ref{alg:EMC}. Briefly, the CPU controls the overall
procedure and streams all required data to the GPU as well as writes
the output. Hence the diffraction patterns are initially loaded and
copied into GPU memory. In each EMC iteration, the compute intensive
steps \eqref{eq:estep}, \eqref{eq:EstepG}, \eqref{eq:cMstep1} and
\eqref{eq:cMstep2} are all evaluated on the GPU.

To discuss those steps, let a `chunk' $C_c$ denote a contiguous set of
rotations, writing $R^c = \{R_{j}; \; j \in C_{c}\}$, $C_{c} := \{j;
\; j_{-}^{c} \le j < j_{+}^{c}\}$. Let the number of rotations in the
set $C_{c}$ be denoted by $|C_c|$, and the total number of chunks as
indexed by $c$ by $\Mchunks$. The computations in the e-step
\eqref{eq:estep} and the M-step \eqref{eq:cMstep1} are partitioned
into chunks naturally. Neither of those steps involve a normalization
over the rotational space, and therefore we can easily calculate and
update \emph{partial} slices $W^c_{ij}$ in a GPU kernel. Each kernel
uses $|C_c|$ blocks, and each block takes care of the computations for
one rotation.

The E-step \eqref{eq:EstepG} and the c-step \eqref{eq:cMstep2} are
more complex due to the normalization over the rotational space. For
the latter kernel $\Mdata \times |C_c|$ blocks are used and each
thread in the block calculates one value of the nominator in
\eqref{eq:EstepG}. The normalization is then performed in a separate
final sweep. In the c-step the updated model is determined by
averaging among chunks, rotations, and diffraction patterns. Here the
numerator and denominator in \eqref{eq:cMstep2} are calculated
separately with the division as a final step. The associated GPU
kernel uses $|C_c|$ blocks, and each block handles one slice in
$R^{c}$.

\subsection{Distributed implementation on a GPU cluster}
\label{subsec:mGPUs}

As argued previously, partitioning the rotational space implies that
the algorithm fits efficiently into memory. This scheme is also simple
enough to be extended to GPU clusters under our preferred master-slave
approach.  Additionally, our distributed EMC algorithm has the feature
of not only dividing the computations of the most expensive step (the
E-step, see Figure~\ref{fig:prof}), but also that it localizes the
computations of the corresponding photon fluence $\phi$, which is the
second most expensive step.

\begin{figure}[!htb] 
\centering
\includegraphics[width=\textwidth]{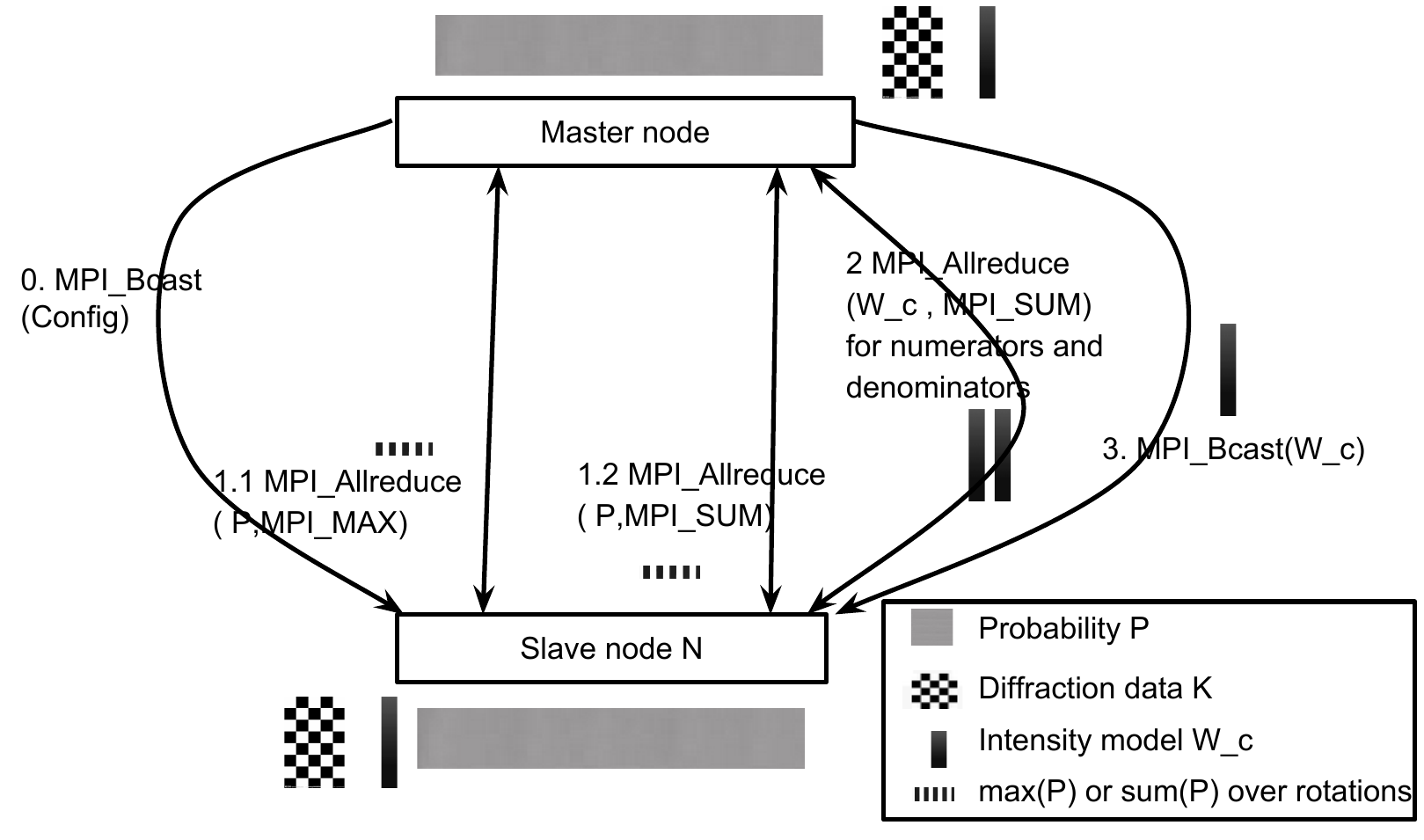} 
\caption{Communication pattern between the nodes for the distributed
  EMC implementation. The rectangles representing data have the
  correct scale with respect to the variable they represent (this is
  the small case of Table~\ref{tab:dataDe}). \new{Step 0 is the
    initialization phase, where only configuration is broadcast, and
    diffraction patterns $K$ and initial model W\_c are fetched by
    each node.} For each EMC iteration, intensity updates are
  performed via steps 1 through 3 among the GPU nodes. W\_c in the
  figure is the intensity model $\Wgrid$.}
\label{fig:art} 
\end{figure}

In the implementation, we associate each GPU with one CPU at the same
cluster node\new{, hence uses one MPI process per GPU, and we
  designate one such CPU/GPU pair as the master node.} For every EMC
iteration, each pair takes care of local computations and synchronizes
when necessary. The master CPU has the overall control of how EMC
synchronizes over the nodes, and each CPU is in charge of the local
GPU computations. Communications between the nodes are thus only
needed in 3 places. Firstly, diffraction patterns and algorithmic
configurations must be broadcast before the algorithm
starts. Secondly, for each chunk $C_c$, local estimated probabilities
$P^c$ must be normalized globally over all nodes. Thirdly and finally,
the local model $\Wgrid^c$ must be merged (averaged) at the end of
each iteration. \new{The data flow among the nodes in our
  implementation is shown in Figure~\ref{fig:art}}. This procedure is
a special case of the fully distributed EMC, which we now proceed to
discuss.

\subsection{Fully distributed EMC}
\label{subsec:fdEMC}

For large enough datasets, the diffraction patterns themselves also
need to be distributed as they no longer fit on a single node. The
chunks $C_c$ are now sets of two-dimensional indices, $C_c :=
\{(j,k);\; k_- \le k < k_+, j_- \le j < j_+ \}$. With this Cartesian
grid-like topology, data is either communicated over the rotational
space (along the $j$-direction), or over image space (along the
$k$-direction).

\begin{figure}[!htb] 
\centering 
\includegraphics[width=\textwidth]{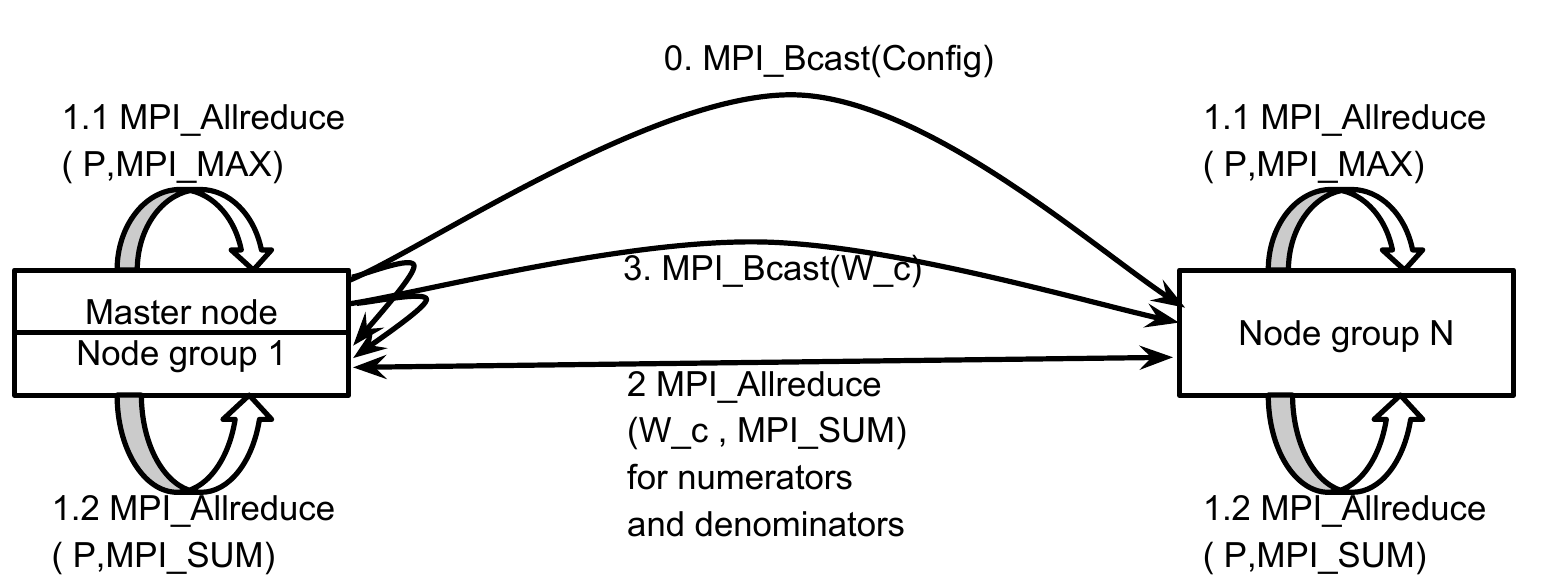}
\caption{Communication pattern of the fully distributed EMC
  implementation. A node group is a set of GPU/CPU pairs which share
  the same partition of diffraction data $K$. Communications among
  node groups are illustrated by thin arrows, and communications
  within a node group are shown in wide arrows. The master node works
  as a normal node in node group 1 with the exception for steps 0 and
  3. According to the notation in the figure, W\_c is the intensity
  model $\Wgrid$. After the configuration is broadcast, every node
  fetches their own portion of $K$, and the same initial model
  $\Wgrid$.}
\label{fig:art_full} 
\end{figure}

The steps that do not involve the diffraction patterns in a global
sense, namely the E-step \eqref{eq:EstepG} and the fluence calculation
of the M-step \eqref{eq:MstepG_S}, are not affected by this novel way
of partitioning data, and can therefore be implemented as previously
described. The remaining steps \eqref{eq:cMstep1}--\eqref{eq:cMstep2}
require data to be broadcast in the $k$-direction. The resulting data
flow among the nodes in our implementation is implicitly shown in
Figure~\ref{fig:art_full}. Step 1.1 and 1.2 in
Figure~\ref{fig:art_full} are only necessary for GPUs that share the
same Cartesian column in $K$, while steps 0, 2, and 3 are global
communications. Further details of the fully distributed EMC are
listed in Algorithm~\ref{alg:EMC_ful}. \new{Note that, in both
  distributed EMC implementations, diffraction patterns $K$ and the
  initial model $\Wgrid^{0}$ are pulled by each node according to the
  data configuration.}

\begin{algorithm}[!htb] 
\begin{algorithmic} 
\State \textbf{Input:} Diffraction patterns $K$, initial guess of the
3D intensity distribution $\Wgrid^{(0)}$.

\State \textbf{Distribute:} Divide all computational nodes into a
Cartesian grid. Distribute diffraction patterns $K$ along the
$k$-direction, and probabilities $P$ along the $j$-direction.
\end{algorithmic}
\begin{algorithmic}[1] 
\Repeat \State At each node copy local data from CPU to GPU;

\State Execute expansion step \eqref{eq:estep}: at each GPU and for
each partition $c$, expand $\Wgrid^{(n)}$ into $W_{ij}^c$, using the
GPU kernel $\Global \void \text{ expansion } \Lkernel |C_c|,
256\Rkernel$;

\State Execute expectation step \eqref{eq:EstepG}: compute the
probabilities $P$ as in \eqref{eq:EstepG} (steps 1.1 and 1.2 in
Figure~\ref{fig:art_full}). The GPU kernel is $ \Global \void \text{
  calculate\_probability} \Lkernel \Mdata \times |C_c|,
256\Rkernel$. Normalization is performed among nodes that share the
same distribution of $K$, hence a $\Reduce$ operation in the
$j$-direction.
    
\State Execute the maximization step \eqref{eq:cMstep1}: update
$W_{ij}^c$ using the GPU kernel $\Global \void \text{ update\_slices}
\Lkernel |C_c|,256 \Rkernel$;

\State Execute the compression step \eqref{eq:cMstep2}: update
$\Wgrid$ according to \eqref{eq:cMstep2}, the numerators and
denominators are calculated via the GPU kernel $\Global \void \text{
  insert\_slices} \Lkernel |C_c|,256 \Rkernel$;

For all nodes, use $\Reduce$ to transfer and add the numerators and
denominators separately, and then perform the final division by the
GPU kernel $\Global \void \text{ insert\_slices\_division} \Lkernel
\lceil \frac{\Mrot}{256} \rceil ,256 \Rkernel$; Finally determine
$\Wgrid^{n+1}$ in the c-step as in \eqref{eq:cMstep2} (step 2 in
Figure~\ref{fig:art_full}).

\State Use $\Bcast$ to update $\Wgrid^{n+1}$ for every node (step 3 in
Figure~\ref{fig:art_full}).

\Until{\mbox{change in either $\Wgrid$ or likelihood is small enough}}
\end{algorithmic} 
\caption{Fully distributed data-parallel version of the EMC algorithm.} 
\label{alg:EMC_ful} 
\end{algorithm}

\new{The data transfers, in both the distributed EMC and in the fully
  distributed EMC implementation, can be implemented
  efficiently. Firstly, the diffraction patterns are fetched by each
  node in the initial phase. Secondly, when normalizing the
  probability, only maximal rotational probabilities and sums are
  necessary, and we may use $\Reduce$. Thirdly, for the fully
  distributed EMC, the use of a Cartesian communicator streamlines the
  communications.}

\subsection{Adaptive EMC}
\label{subsec:adaptiveEMC}

Previously, we aimed for an efficient execution by distributing data
and using an increasing number of compute nodes. In most actual
experience with the method a fairly highly resolved rotational space
is used and one typically observes a likelihood which slowly but
steadily improves. The performance critical steps of the method, the
E-step \eqref{eq:EstepG} and part of the combined M/c-step
\eqref{eq:cMstep2}, scale directly with the discretization of the
rotational space $\Mrot$. Hence it seems reasonable not to waste
computational time using a large value of $\Mrot$ before the
likelihood has improved sufficiently. A simple way to achieve this is
to start using some small value of $\Mrot$, $\Mrot(d_{0})$ say, and
increase $d$ in \eqref{eq:Mrot} whenever the current value seems too
small to improve the likelihood any further. In practice we increase
$d$ by 1 when
\begin{align} 
  \label{eq:diff_likelihood}
  \left\lvert L^n-L^{n-1} \right\rvert \leq 0.01 \left \lvert L^n \right\rvert, 
\end{align}
with $L^{n}$ the likelihood at iteration $n$.


\section{Experiments}
\label{sec:experiments}

We now proceed to investigate the achievable performance of our
implementation. Our datasets are either real ones and of similar
quality to those currently being processed, or are synthetic and of
considerably larger size to be able to assess future performance
profiles. In particular, we will explore the possibility to obtain a
highly resolved image from a very large diffraction dataset on a
cluster consisting of 100 GPUs.

\subsection{Setup and basic profiling}
\label{sec:setup}
\new{We ran our experiments using a 32-node homogeneous GPU
  cluster. Each node of the cluster is equipped with 4 six-core Intel
  Xeon E5-2620 CPUs and 4 Nvidia GeForce GTX 680 GPUs. All the CPU
  cores operate at 2.0 GHz with 32K L1i and L1d catches. They are
  organized as 2 NUMA nodes with a total of 64 GB memory. Each GTX 680
  GPU has 4GB memory, and the reported nominal peak performance in
  single precision of matrix multiplication, matrix left division, and
  the Fast Fourier Transform are $[966, 573, 101]$ GFLOPS,
  respectively \cite{GPUbench}. Cluster nodes are interconnected using
  a QDR Infiniband with a bandwidth of 32 Gbit/s. We measured the
  bandwidth of CPU/GPU connections by doing a host-to-device and
  device-to-host memory copy, and we found that the bandwidth in both
  direction is around 3 GB/s. For the compilers and libraries, we used
  GCC~4.4.6, CUDA~5.0, and Open MPI~1.5.4. By considering the
  inter-node MPI bandwidth, the intra-node CPU/GPU bandwidth and the
  data volume which needs to be transferred, we judge that our
  implementation is not bandwidth bound.}

\begin{figure}[!htb]
  \subfloat{\includegraphics[width=0.33\textwidth]{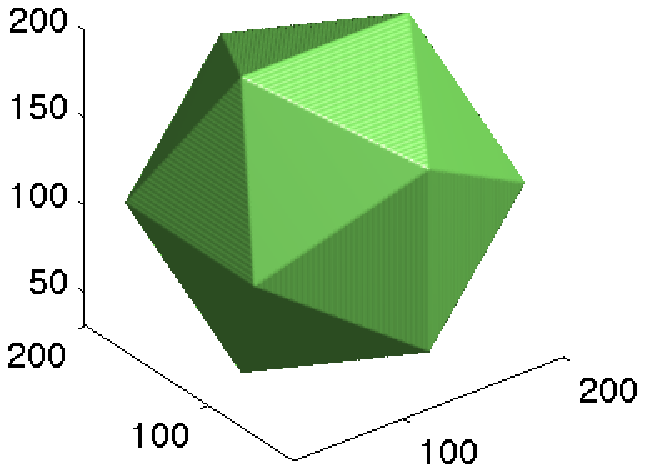}}
  \subfloat{\includegraphics[width=0.33\textwidth]{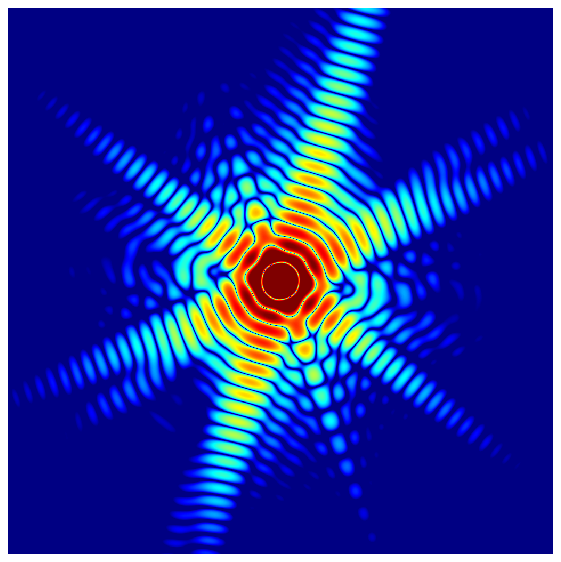}}
  \subfloat{\includegraphics[width=0.33\textwidth]{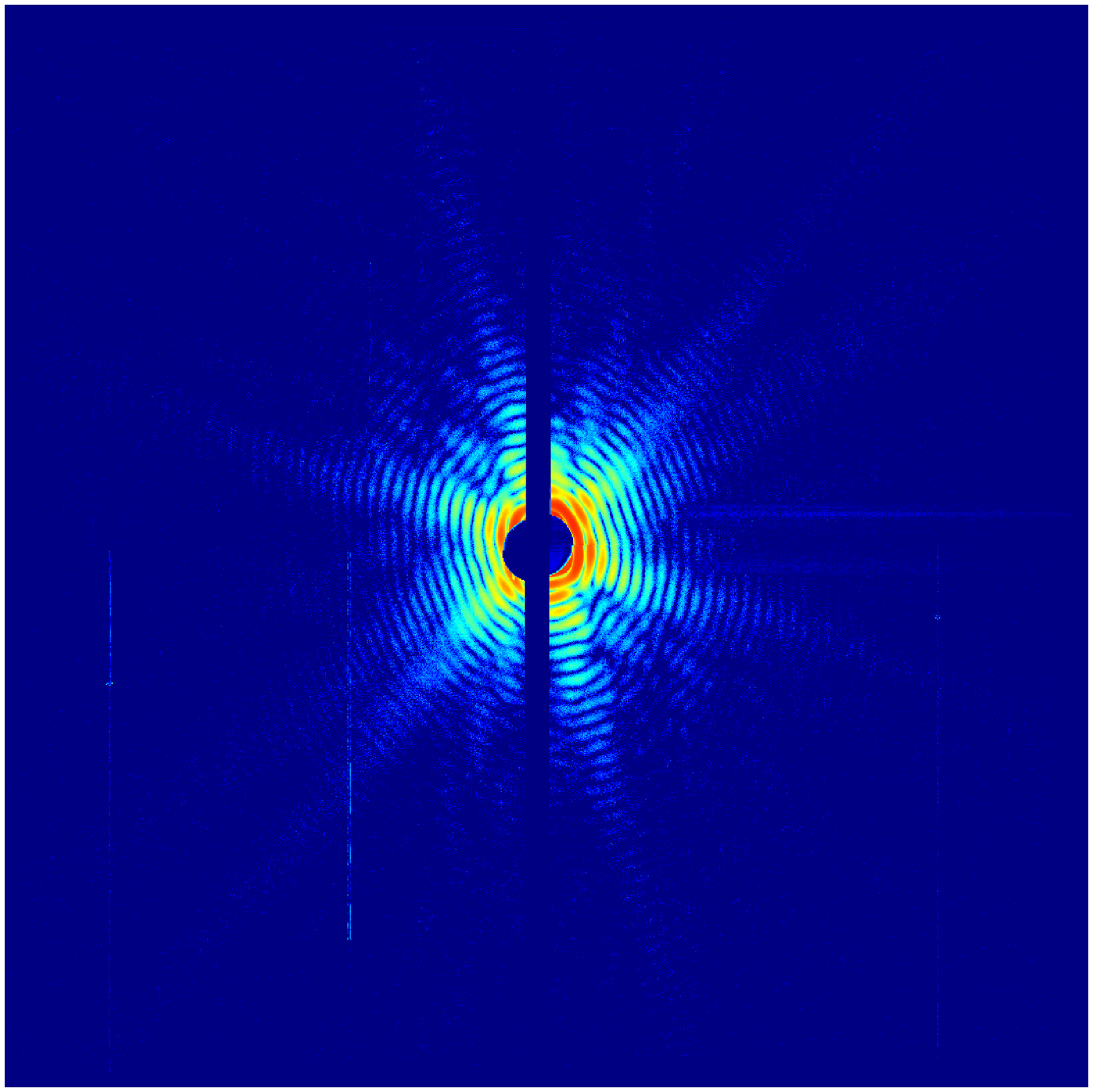}}
  \caption{\textit{Left:} real-space icosahedron, \textit{middle:}
    resulting synthetic diffraction pattern, \textit{right:} X-ray
    diffraction pattern from the Mimivirus. The two diffraction
    patterns are plotted in logarithmic scale.}
\label{fig:dataset}
\end{figure}

We used mainly two different diffraction datasets. A smaller case
consisting of 198 images from an X-ray diffraction experiment with the
giant Mimivirus \cite{TomasPhD,mimivirus_Xrays} performed at the Linac
Coherent Light Source (LCLS). A larger case was obtained through
synthetic simulation for an icosahedral shape and consists of 1000
frames. Figure~\ref{fig:dataset} displays samples of these diffraction
patterns.

In Table~\ref{tab:dataDe} we list the sizes of all relevant data in
our experiments. The experiments were configured to reconstruct a
small ($64 \times 64 \times 64$), or, respectively, a large ($128
\times 128 \times 128$) 3D intensity model in single precision. In
\S\ref{subsec:efficiency} we also ran a `giant' case consisting of our
synthetic dataset, but duplicated ten times (i.e. a total of 10,000
frames).

\begin{table}[!htb]
  \centering
  \begin{tabular}{rrrr}
    \hline  Set $\#$ &data &$\Mdata$& $\Mpix$ \\ \hline 
    1 & Mimivirus & 198 & 4096 \\
    2 & Mimivirus & 198 & 16384 \\
    3 & synthetic & 1000& 4096 \\
    4 & synthetic & 1000 & 16384 \\
    5 & synthetic & 10000 & 4096 \\
    6 & synthetic & 10000 & 16384 \\
    \hline
  \end{tabular}
  \caption{Sizes of the different datasets used in our
    experiments. Note that the value of $\Mpix$ is the result after
    binning the raw data $1024 \times 1024$ into a coarser $64 \times
    64$ (or $128 \times 128$) format, and note also the relation
    $\Mgrid = \Mpix^{3/2}$. In all these experiments we used the value
    $\Mrot = 86520$.}
  \label{tab:dataDe} 
 \end{table}

\begin{figure}[htb] 
  \centering 
  \includegraphics{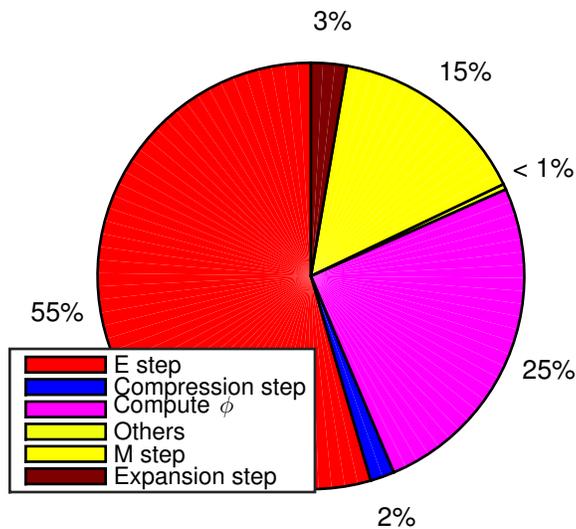} 
  \caption{Profiling the single GPU EMC implementation. The two most
    expensive operations are the E-step \eqref{eq:EstepG} and updating
    the photon fluence \eqref{eq:MstepG_S}.}
  \label{fig:prof}
\end{figure}

The result of profiling our single-GPU implementation provided a
motivation for our approach to distribute data in our multi-GPU
implementation. We profiled the single-GPU implementation by
reconstructing at low resolution the Mimivirus dataset (Set \#1 in
Table~\ref{tab:dataDe}) on one \new{Nvidia GeForce GTX
  680}. Figure~\ref{fig:prof} displays the computational statistics
after averaging over the first 10 iterations. As expected, the E-step
was the most expensive step, consuming more than 55\% of the total
time. This means that it is preferable to distribute data along the
rotational space (along the $j$-direction), at least given this
relatively small number of diffraction patterns. Since the probability
$P_{jk}$ has the same size as the photon fluence $\phi_{jk}$, this
also means that the second most expensive step, i.e. the computation
of $\phi$, parallelizes very well.

\subsection{Performance analysis}
\label{subsec:efficiency}

In this section we report results from our distributed EMC from
\S\ref{subsec:mGPUs} for both the Mimivirus dataset and the synthetic
dataset on up to 32 GPUs. We also look at the performance of the fully
distributed EMC from \S\ref{subsec:fdEMC} using the `giant' synthetic
dataset, consisting of 10,000 diffraction patterns. In contrast to
these large-scale experiments, we will also explore the performance of
adaptive EMC on a single GPU.

\begin{figure}[!htb] 
  \centering
  \includegraphics{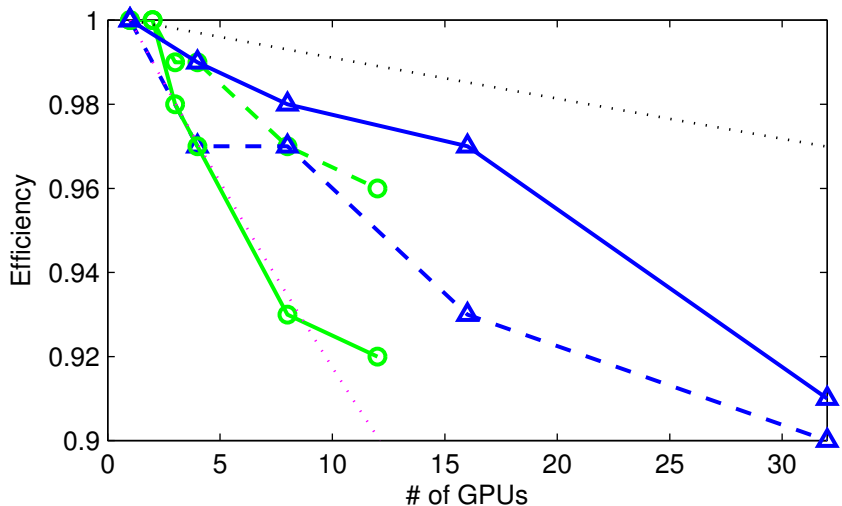} 
  \caption{Measured efficiencies for dataset \#1--4.
    \textit{Circles:} real Mimivirus data, \textit{triangles:}
    synthetic dataset, \textit{solid:} $\Mpix =16384$,
    \textit{dashed:} $\Mpix = 4096$.  \textit{Upper dotted line:}
    Amdahl's efficiency \eqref{eq:pre_eff} with $B=0.001$,
    \textit{lower dotted line:} $B=0.01$.}
  \label{fig:eff} 
\end{figure}

We first ran some experiments with the distributed EMC implementation
as discussed in \S\ref{subsec:mGPUs}. We measured Amdahl's efficiency,
\begin{align} 
  \label{eq:pre_eff} 
  E = \dfrac{T(1)}{nT(n)} =
  \dfrac{T(1)}{T(1)(B+\frac{1}{n}(1-B))} = \dfrac{1}{nB + (1-B)},
\end{align} 
where $n$ is the number of GPUs, and $B$ the fraction of the algorithm
that is serial. The results are shown in Figure~\ref{fig:eff}. As can
be seen we obtain a nearly perfect efficiency, at least up to 32 GPUs.

We next ran some really large-scale experiments for the fully
distributed EMC implementation. In an attempt to follow the topology
of our cluster, we distributed the rotational space (i.e. the
$j$-direction) over a total of 25 nodes, and for the 4 GPUs belonging
to the same node, we distributed the 10,000 diffraction patterns
uniformly.

Defining one multiplication as 1 FLOP and one division as 8 FLOPs, we
find by inspection that one EMC iteration requires about
$\Mrot\times\Mdata\times\Mpix \times 27 \times 10^{-9}$
GFLOPs. Table~\ref{tab:flops} lists the average execution time per
iteration and the achieved GFLOPS per GPU. It is remarkable that we
loose less than 4\% floating point performance at 100 GPUs compared to
16 GPUs. In fact, the fully distributed EMC implementation achieves a
higher floating point performance when compared to the single GPU
implementation (32.9 GFLOPS and 39.4 GFLOPS, respectively, for $\Mpix
= 4096$ and $\Mpix = 16384$). Indeed, these figures compare favorably
with the online GPU benchmark \cite{GPUbench}, where square
matrix-matrix multiplication in single precision achieves
$[4.4,32.6,181]$ GFLOPS, respectively, at the comparable matrix sizes
$N = [16384,65536,262144]$.

\begin{table}[!htb] 
  \centering \begin{tabular}{rrrrr} 
    \hline
    & \multicolumn{2}{r}{$\Mpix = 4096$} & 
    \multicolumn{2} {r}{$\Mpix = 16384$} \\
    \# GPUs & Time (s) & GFLOPS/GPU & Time (s) & GFLOPS/GPU \\
    \hline 
    16 & 164.6 &$36.3$& 552.2 & $43.3$\\ 
    32 & 83.5& $35.8$ & 281.2& $42.5$ \\ 
    64 & 42.3& $35.3$&141.6 & $42.3$\\ 
    96 & 28.3 & $35.2$& 95.4& $41.8$\\ 
    100 & 27.2 & $35.2$& 91.6 & $41.8$\\
    \hline 
  \end{tabular} 
  \caption{Average execution time and floating point performance per
    GPU and per iteration using the fully distributed EMC.}
  \label{tab:flops} 
\end{table}

	

\begin{figure}[!htb]
 \centering 
 \includegraphics[width=\textwidth]{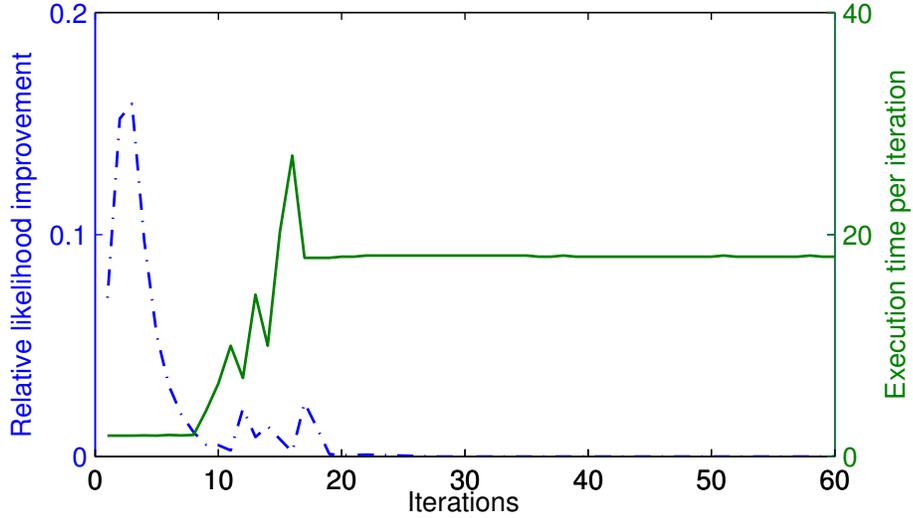}
 \caption{Difference of likelihood \eqref{eq:diff_likelihood} and the
   corresponding execution time (in seconds) for the first 60
   iterations.}
\label{fig:dyn} 
\end{figure}

Finally, we also performed some experiments with our adaptive EMC
algorithm. For simplicity we used a single GPU only, and we
reconstructed a small $64\times 64\times 64$ intensity model. For the
adaptivity we increased $d$ in \eqref{eq:Mrot} from 5 to 12 according
to the likelihood-based criterion
\eqref{eq:diff_likelihood}. Figure~\ref{fig:dyn} displays the relative
difference of likelihood together with the execution time per
iteration. As expected, the execution time increases as the resolution
of the rotational space increases. Whenever $d$ is increased, there is
a sharp peak in the relative likelihood difference, indicating
iterations that successfully increase the likelihood. The performance
gain for the adaptive version is quite remarkable, $7341.6$ seconds
compared to $14524$ seconds for the original version, using 60
iterations for both runs. By the very simplicity of this approach, we
expect that this gain of a factor of about 2 remains also for larger
load cases.

\subsection{Scalability} 
\label{subsec:sca}

\new{Instead of a scaling for a fixed problem-size, we study a metric
  of scalability that takes the data volume into account.}  Define the
effective problem size as $V(C) \equiv \Mrot(C) \times \Mpix(C) \times
\Mdata(C)$ for $C$ a problem configuration. For two different such
configurations $C_{1}$ and $C_{2}$, the scalability $S(C_1,C_2)$ is
then defined as
\begin{align}
  \label{eq:scale} 
  S(C_1,C_2) &= \dfrac{T(C1)/T(C2)}{V(C_1)/V(C_2)}, 
\end{align}
where $T(C)$ is the execution time for configuration $C$. \new{A
  superlinear scalability $S(C_1,C_2) < 1$ means that the program
  works more efficiently with configuration $C_2$, and also suggests
  that the program does not make full use of the computing power in
  $C_1$. The situation $S(C_1,C_2) > 1$ may happen for a problem which
  is compute-bound.}

\begin{figure}[!htb] \centering
\includegraphics[width=\textwidth]{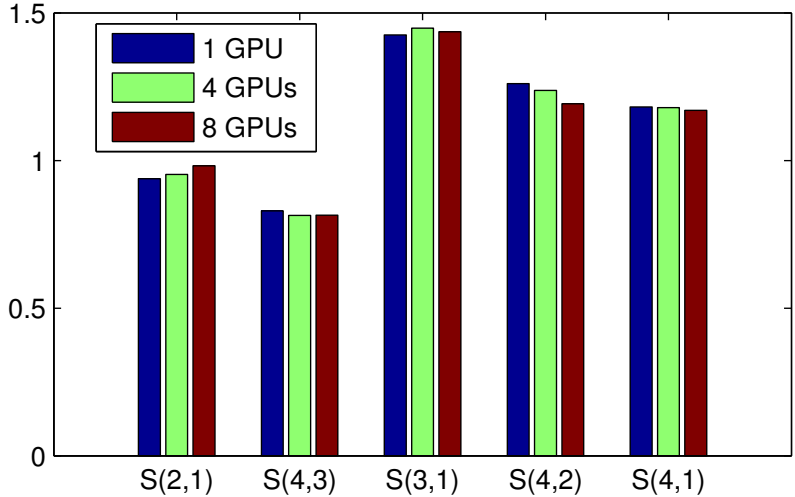}
\caption{The scalability $S(C_1,C_2)$ of configuration $C_1$ and
  $C_2$, according to \eqref{eq:scale}, with $C_1$ and $C_2$ chosen
  among cases 1--4 in Table~\ref{tab:dataDe}.}
\label{fig:scale} 
\end{figure}

Figure~\ref{fig:scale} displays the scalabilities of different
configurations. Notably, the cases $S(2,1)$ and $S(4,3)$ show
superlinear scalability. Since only $\Mpix$ is changed in both cases,
we judge that this is an effect of that the GPU kernels of the c- and
e-step are not fully loaded. To the contrary, increasing $\Mdata$
makes the scalabilities $S(3,1)$ and $S(4,2)$ larger than 1. For the
latter case we see that as we add more GPUs, we obtain a slightly
better scalability. This indicates that as the size of the datasets
increases, the fully distributed EMC becomes a favorable choice.

Finally, with $S(4,1)$, we compare a synthetic dataset to a real one,
and we simultaneously increase $\Mpix$ and $\Mdata$. However, the
values in this group only slightly differ from eg.~$S(4,2)$, which
suggests that $\Mdata$ plays a more prominent role than $\Mpix$ when
measuring performance.

\section{Conclusions}
\label{sec:conclusions}

We have implemented the EMC algorithm for the assembly of randomly
oriented diffraction patterns on GPUs using the CUDA framework and
have extended the algorithm to run efficiently on multiple GPUs. We
use two different partition schemes depending on the number of
diffraction patterns. For a medium-sized dataset we partition the
rotational space only, while for a large amount of data we also
distribute the images themselves. We observe almost linear speedups
for up to 100 GPUs and a parallel efficiency that thus compares very
well with other MPI/CUDA applications
\cite{jacobsen2010mpi,wang2013cpu,zaspel2012solving}. We also devised
an adaptive technique by which the resolution is increased on par with
the increase in likelihood. In our experiments this idea worked very
well and we expect similar ideas to be useful in implementations on
site where data is processed in a streaming fashion
\cite{automatic_id}.

It seems likely that our implementation is going to develop and adapt
further as larger datasets become available. Hopefully, the present
software framework can substantially shorten the development cycle for
novel algorithms targeting large and noisy datasets.

Hit-ratios and data quality at the LCLS is steadily improving and in
2016 the European XFEL will become operational and provide a
repetition rate of 27,000 Hz, a 200-fold increase compared to the
LCLS. With this, data analysis will become a bottleneck and EMC in
particular is the main computational step in this analysis. With this
work we hope to ensure that data analysis can keep up with the rapid
development of X-ray laser facilities, while simultaneously enabling
the study of biological particles from single proteins to viruses.

	
\section*{Acknowledgment}

This work was financially supported by by the Swedish Research Council
within the UPMARC Linnaeus center of Excellence (S.~Engblom, J.~Liu)
and by the Swedish Research Council, the Knut och Alice Wallenberg
Foundation, the European Research Council, the Röntgen-Ångström
Cluster, and the Swedish Foundation for Strategic Research
(T.~Ekeberg, J.~Liu).

Input and suggestions on an earlier draft of the paper from Filipe
R.~N.~C.~Maia and Janos Hajdu are hereby gratefully acknowledged.
	


	
\newcommand{\doi}[1]{\href{http://dx.doi.org/#1}{doi:#1}}
\newcommand{\available}[1]{Available at \url{#1}}
\newcommand{\availablet}[2]{Available at \href{#1}{#2}}
	
\bibliographystyle{abbrvnat} \bibliography{eel}	
	\end{document}